\documentclass[aps,prd,showpacs,floats,floatfix,twocolumn,nofootinbib,amssymb,amsmath]{revtex4}
\usepackage{amsmath,amsfonts}
\usepackage{hyperref}
\usepackage{graphicx}
\usepackage{bm}
\usepackage{units}

\renewcommand{\Re}{\mathrm{Re}}

\newcommand{\beq}{\begin{equation}}
\newcommand{\eeq}{\end{equation}}
\newcommand{\beqa}{\begin{eqnarray}}
\newcommand{\eeqa}{\end{eqnarray}}

\newcommand{\vs}{{\hat s}}
\newcommand{\vh}{{\hat h}}
\newcommand{\bra}{\langle}
\newcommand{\ket}{\rangle}
\newcommand{\FF}{\mathcal{F}}
\newcommand{\Msun}{M_{\odot}}
\newcommand{\ud}{\,\mathrm{d}}
\newcommand{\baln}{\begin{align}}
\newcommand{\ealn}{\end{align}}
\newcommand{\ecc}{eccentricity }
\newcommand{\xmod}{\ensuremath{x}-model }
\newcommand{\bs}{\begin{subequations}}
\newcommand{\es}{\end{subequations}} 
\def\no{\nonumber \\}

\begin{document}

\title{The Effect of Eccentricity on Searches for Gravitational-Waves from Coalescing Compact Binaries in Ground-based Detectors} 

\author{Duncan A. Brown${}^1$, Peter J. Zimmerman${}^1$}

\affiliation{${}^1$ Department of Physics, Syracuse University,
Syracuse, NY 13244}

\date{\today}

\begin{abstract}
Inspiralling compact binaries are expected to circularize before their
gravitational-wave signals reach the sensitive frequency band of ground-based
detectors. Current searches for gravitational waves from compact binaries
using the LIGO and Virgo detectors therefore use circular templates to
construct matched filters. Binary formation models have been proposed which
suggest that some systems detectable by the LIGO--Virgo network may have
non-negligible eccentricity.  We investigate the ability of the restricted 3.5
post-Newtonian order TaylorF2 template bank, used by LIGO and Virgo to search
for gravitational waves from compact binaries with masses $M \le 35 M_\odot$, to detect
binaries with non-zero eccentricity. We model the gravitational waves from
eccentric binaries using the $x$-model post-Newtonian formalism proposed by
Hinder \emph{et al.} [I. Hinder, F. Hermann, P. Laguna, and D.  Shoemaker,
arXiv:0806.1037v1]. We find that small residual eccentricities ($e_0 \lesssim
0.05$ at $40$~Hz) do not significantly affect the ability of current LIGO
searches to detect gravitational waves from coalescing compact binaries with
total mass $2 M_\odot < M < 15 M_\odot$. For eccentricities $e_0 \gtrsim 0.1$,
the loss in matched filter signal-to-noise ratio due to eccentricity can be
significant and so templates which include eccentric effects will be required
to perform optimal searches for such systems.
\end{abstract}


\maketitle

\section{Introduction}
\label{s:Introduction}  

In October 2007 the Laser Interferometer Gravitational-Wave Observatory (LIGO)~\cite{Barish:1999vh} completed its fifth science run,
collecting one year of coincident data at design sensitivity~\cite{Abbott:2007kva}. After a period of detector upgrades~\cite{Smith:2009bx}, LIGO and the
French-Italian Virgo detector~\cite{Acernese:2005yh} began
a new observing campaign in July 2009 and construction of the Advanced LIGO
(AdvLIGO) detectors~\cite{Fritschel:2003qw} is underway. 
The inspiral and coalescence of compact binary systems consisting of neutron
stars (NS) and/or black holes (BH) is a promising source of gravitational waves for the LIGO--Virgo network~\cite{Abbott:2009tt, Abbott:2009qj}.  
Optimal searches for inspiral signals use the method of
matched-filtering~\cite{SignalRefNo1,SignalRefNo2}, 
in which the time-series of the detector is cross-correlated against a
set of theoretical template waveforms. The utility of this technique rests on
how well the template waveforms model the signal being sought.
Significant progress has been made in modeling gravitational waves from
compact binaries using post-Newtonian (PN) theory~\cite{Blanchet:2006zz} and numerical
relativity (NR)~\cite{Hannam:2009rd}.
For binaries that evolved through typical main sequence evolution~\cite{Kalogera:2004tn,Kalogera:2004nt},
radiation reaction will cause them to circularize by the time the
frequency of their gravitational waves enters the sensitive frequency band of
ground-based detectors such as LIGO and Virgo~\cite{Peters:1964zz, Peters:1963ux}.  
Alternative formation mechanisms have been proposed that may yield binaries with
non-negligible eccentricity in the LIGO band~\cite{O'Leary:2008xt,Wen:2002km}. 
The scattering of stellar mass BHs in galactic cores containing a
super-massive BH may lead to binary formations with high
eccentricities; $\sim 90\%$ with $e_0 > 0.9$, where $e_0$ denotes the eccentricity of the binary when it enters the sensitive band of the detector~\cite{O'Leary:2008xt}. 
In such an encounter, the two BHs radiate enough energy to form a
bound system, but do not fully radiate their eccentricity in GWs until they coalesce.
Estimates suggest that the rate of such BH binary coalescences detectable by AdvLIGO may be as high as 100 yr$^{-1}$~\cite{O'Leary:2008xt}. 
Binary-binary interactions in the central regions of
globular clusters may also be a source of gravitational waves with non-zero
eccentricities. Multi-body interactions in globular clusters may result in the
formation of a stable hierarchical triple. If the orbital planes
of the inner and outer binary with respect to the center of mass are
highly inclined with respect to one another, Kozai resonance increases
the eccentricity of the inner binary~\cite{Wen:2002km}. It has been estimated that 
$\sim 30\%$ of binaries formed in this way will have eccentricities $e_0 >0.1$ when they
enter the AdvLIGO frequency band ($\sim 10$ Hz). 
However, most template families, including those used by LIGO to search for
inspiral signals, make use of the \emph{quasi-circular} approximation. In this
scheme, the binary system evolves secularly through a sequence of
circular orbits~\cite{Damour:2000zb, Blanchet:2006zz}; 
eccentricity is neglected. 

In recent years there has been significant progress in modeling
eccentric binaries using post-Newtonian theory.
These calculations use a combination of multi-scale methods and variation of
constants~\cite{Gopakumar:1997bs, Damour:2004bz, Konigsdorffer:2006zt}
to include conservative post-Newtonian effects in the
phase evolution~\cite{Damour:2004bz} in addition to the effect of radiation
reaction. The conservative dynamics
account for small oscillations in phase evolution which occur on
time-scales of the order of an orbital period.
At present, the 3~PN conservative contributions to the phase evolution have
been derived in the quasi-Keplerian parameterization
~\cite{Memmesheimer:2004cv} and the 3~PN reactive contribution is
expected shortly now that the calculation of the 3~PN
flux~\cite{Arun:2007sg} is complete. 
  
The first investigation of the effect of eccentricity on searches for
gravitational waves 
was performed by Martel and Poisson~\cite{Martel:1999tm}. In their study,
they examined the effect of the
leading-order radiation effects~\cite{Peters:1964zz}
and concluded that circular templates
were effective at capturing sources with small residual eccentricity.
This problem was recently revisited by Cokelaer and Pathak~\cite{Cokelaer:2009hj}
using a discrete template bank in an effort to more accurately model an
inspiral search.  The qualitative conclusions of Ref.~\cite{Martel:1999tm} 
were upheld by Ref.~\cite{Cokelaer:2009hj}, despite small quantitative differences due to
a difference in the numerical precision used in the computations
~\cite{Cokelaer:2009hj}.
These studies used waveforms computed at leading order; both
conservative effects on the waveform evolution and higher order post-Newtonian
corrections to the phase evolution were omitted. Recent work by Tessmer and
Gopakumar, based on templates which include conservative post-Newtonian effects,
suggests that circular templates may not be adequate to detect systems
with residual eccentricity~\cite{Tessmer:2007jg}.

In this paper, we investigate whether the waveforms used in current LIGO
inspiral searches are adequate to \emph{detect} gravitational waves from compact
binaries with residual eccentricity. To model eccentric signals, we use
the post-Newtonian eccentric waveform family proposed by Hinder \emph{et
al.}~\cite{Hinder:2008kv}, which has been calibrated against numerical
simulations of eccentric BH binaries. In Sec.~\ref{s:PostNewtonian} we review
the waveforms used as eccentric signals and in Sec.~\ref{s:DataAnalysis} we
review the data analysis techniques used. Sec.~\ref{s:Results} presents
our results for the current and AdvLIGO detectors.

\section{Eccentric Post-Newtonian Waveforms}
\label{s:PostNewtonian}
The effect of gravitational radiation on the orbital motion of eccentric
binaries was first calculated by Peters and Mathews~\cite{Peters:1964zz,
Peters:1963ux}.  Post-Newtonian
corrections to the Peters and Mathews result were first calculated in Ref.~\cite{Gopakumar:1997bs} and
later refined using an improved method of variation of constants in
Ref.~\cite{Damour:2004bz}.  The PN equations of motion in
Ref.~\cite{Damour:2004bz} were written in terms
of the mean motion $n$ and time eccentricity $e_t$ and the 
3.5~PN phase evolution has been calculated using these variables~\cite{Konigsdorffer:2006zt}.
The recent progress in the numerical evolution of BH binaries now allows
post-Newtonian models to be calibrated against the gravitational waves
extracted from simulations. Hinder \emph{et. al.} investigated the agreement
of eccentric post-Newtonian waveforms with the gravitational waves extracted
from an equal mass BH binary with eccentricity $e_0 = 0.1$~\cite{Hinder:2008kv}. Based on this study
they proposed a modification of the eccentric PN waveforms (called the
$x$-model) in which 
the dynamical quantities are written in terms of a variable related to the
orbital frequency $x=(M\omega)^{2/3}$ and the time
eccentricity $e_t$. This waveform gives better agreement to the numerical data
than the waveform written in terms of $n$ and $e_t$ and so we use it to model
gravitational waves from eccentric binaries in our analysis.
In the remainder of this section, we review the $x$-model formalism introduced 
in Ref.~\cite{Hinder:2008kv}. 

\subsection{Conservative post-Newtonian Dynamics}
\label{s:ConservativeDynamics}
The post-Newtonian model of Ref.~\cite{Hinder:2008kv} is more easily understood if we first
consider the evolution of two point masses $m_1$ and $m_2$ in the purely
Newtonian case.  In the Newtonian case, conservation of energy $E$ and
angular momentum $J$ dictate the evolution of the orbital elements.
We define the \emph{mean motion} $n$ in terms of the orbital period $P$, $n =
\frac{2 \pi}{P}$.  For Newtonian trajectories on an ellipse ($0<e_0<1$), $n$ is
simply $\sqrt{M/a^3}$ where $M=m_1+m_2$ and $a$ is the semi-major
axis\footnote{We work in units with \(G=c=1\).}.  To parameterize the
equations we introduce the \emph{eccentric anomaly}
$u$. Written in terms of $u$, the relative orbital separation $r$ and the angular
frequency $\dot{\phi}$ take the form
\begin{eqnarray}
  r &=& a \left[1 - e \cos u \right], \\
  \dot{\phi} &=& \frac{n\sqrt{1-e^2}}{{\left[1-e \cos u \right]}^2}.
\end{eqnarray}
To complete the dynamical system we relate the eccentric anomaly to the
mean motion via the Classical Kepler Equation 
\beq
\label{eq:ClassicalKepler} 
  l = u - e \sin u, 
\eeq
where the \emph{mean anomaly} $l$ is given
by integrating $dl = n\, dt$.
In the absence of radiation reaction, $n$ is time-independent and the mean
anomaly is simply \( l = n\,(t - t_0) \).  
These equations form the Keplerian
parameterization of a Newtonian orbit. Given initial conditions \(
\phi_0 \equiv \phi(t_0) \) and \( l_0 \equiv l(t_0) \) along with
numerical values for $e_0$ and $u$ we can compute the trajectories of the
two particles at any time \(t > t_0 \) by root finding for $u$.  

The post-Newtonian dynamics of a system are conservative if they
the respect energy and angular momentum conservation\footnote{This
definition fails at 3.5~PN due to center of mass ``recoil''. A more
rigorous definition is given in Ref.~\cite{Damour:2004bz}.}.  Using the
relations given in Ref.~\cite{Memmesheimer:2004cv} for $n$ and $e_t$ in
terms of $E$ and $J$, one can derive 3~PN quasi-Keplerian equations of
motion by introducing three eccentricities, $e_t,\,e_r$ and $e_{\phi}$,
which account for the variations in the $t$, $r$, and $\phi$ directions. The addition of
higher order conservative PN corrections lead to periastron precession.
We let $\Delta \phi$ represent the angle of precession
during one period $P$. The angle swept out over consecutive periastron
passages is defined 
\beq
  \omega = \frac{2\pi + \Delta\phi}{P} 
\eeq
~\cite{Arun:2007rg, Arun:2007sg, Hinder:2008kv}. We note that \( \omega
\) is a constant in the absence of radiation reaction.

The evolution equations in the conservative dynamics take the
abbreviated form~\cite{Hinder:2008kv}
\beqa
\label{eq:PnSeparation}
  \frac{r}{M} &=& \left(1-e_t \cos u \right)x^{-1} 
    + r_{1\mathrm{PN}} + r_{2\mathrm{PN}}x \no
    & + & r_{3\mathrm{PN}} x^2 + \mathcal{O}(x^{3}), \\
  M\dot{\phi} &=& \frac{\sqrt{1-e_t^2}}{{\left(1-e_t \cos u \right)}^2}x^{3/2} 
    + \dot{\phi}_{1\mathrm{PN}}x^{5/2} + \dot{\phi}_{2\mathrm{PN}} x^{7/2} \no
    & + & \dot{ \phi }_{3 \mathrm{PN}} x^{9/2} + \mathcal{O}(x^{11/2}), \\
\label{eq:PnKepler}
  l &=& u - e_t \sin u + l_{2\mathrm{PN}}x^2 +
          l_{ 3\mathrm{PN} } x^3 + \mathcal{O}(x^{4}),  \\
\label{eq:MeanMotion}
  M\dot{l} & = &  Mn = x^{3/2} + n_{1\mathrm{PN}}x^{5/2} +
    n_{2\mathrm{PN}}x^{7/2} \no 
    & + & n_{3\mathrm{PN}}x^{9/2} + \mathcal{O}(x^{11/2})\ .
\eeqa
Detailed expressions for the PN coefficients are given in the appendix
of Ref.~\cite{Hinder:2008kv}. We note that the PN coefficients
$r_{\mathrm{PN}},\,\dot{\phi}_{\mathrm{PN}}, \ldots $ are functions of both $e_t$
and $u$, whereas $n$ depends on $e_t$ alone.  
Conservative PN trajectories are obtained by first integrating
Eq.~\eqref{eq:MeanMotion} for $l$, and then numerically root solving Eq.~\eqref{eq:PnKepler} 
for $u$ at each time step $t$.  

\subsection{Radiative post-Newtonian Dynamics}
\label{s:ReactiveDynamics}
Radiative post-Newtonian dynamics are needed to describe
evolutions over time-scales in which angular momentum
and energy are carried away from the system by gravitational radiation.
The time dependence of $E$ and $J$ implies that $x$ and $e_t$ are
no longer integrals of motion. The time variation of $n$ and $e_t$
leads to a secular evolution of the orbital parameters.
To describe the secular evolution of the
orbital elements $\phi,\,\dot{\phi},\,r,\,\dot r$, and $l$,
the equations of motion for $x$ and $e_t$
must be extended into the non-conservative regime.  
The 2~PN equations describing the radiative dynamics read\footnote{It is conventional
in the gravitational-wave literature to define the radiative PN order relative
the order at which radiation reaction occurs.}   
\beqa
\label{eq:Mdxdt}
  M\dot{x} & = & \frac{ 2 \eta }{ 15 \left( 1 - e_t^2 \right)^{7/2} } 
  \left( 96 + 292 e_t^2 + 37e_t^4 \right) x^5 
  + \dot{x}_{\mathrm{1PN}} x^6 \nonumber \\
  & + & \dot{x}_{\mathrm{1.5PN}} x^{13/2}
  + \dot{x}_{\mathrm{2PN}} x^7 + \mathcal{O}(x^{15/2}),  \\
\label{eq:Mdedt}
  M\dot{e_t} & = & \frac{ -e_t \eta }{ 15 \left( 1 - e_t^2 \right)^{7/2} } 
  \left( 304 + 121 e_t^2 \right) x^4 
  + \dot{e}_{\mathrm{1PN}} x^5  \nonumber \\
  & + & \dot{e}_{\mathrm{1.5PN}} x^{11/2} + \dot{e}_{\mathrm{2PN}} x^{6}
  + \mathcal{O}(x^{13/2}), 
\eeqa
where $\eta \equiv \mu / M = m_1 m_2 / (m_1+m_2)^2$ is the symmetric
mass ratio.
Again, we refer to the appendix of Ref.~\cite{Hinder:2008kv}
where the PN coefficients are written out explicitly. 
In the adiabatic approximation, the PN coefficients are independent
of $u$, and are solved independently of the Kepler Equation. We
use a fourth order Runge-Kutta-Fehlberg (RKF-45) with adaptive step-size
control~\cite{GSL} to
numerically solve for the radiative dynamics at each time $t >
t_0$. Once $x(t)$ and $e_t(t)$ are obtained, we substitute their values
into Eq.~\eqref{eq:MeanMotion} and solve the ordinary differential equation
(ODE) for $l(t)$. The value of $l(t)$ is
then equated to the RHS of Eq.~\eqref{eq:PnKepler}, giving us our PN
Kepler Equation. We then solve the PN Kepler Equation using
the Mikkola root finding method \cite{1987CeMec..40..329M}. 
The values of $x$, $e_t$, and $u$ are
then substituted into the conservative equations for $r$ and
$\dot{\phi}$. The value of $\dot{r}$  is obtained numerically using a
five-point stencil method and the phase $\phi$ is numerically integrated
using the RKF-45. Repeating this process, we obtain the dynamics
$(r,\, \dot{r},\, \phi,\, \dot{\phi})$ at each time $t>t_0$.

\subsection{Formulation of the Dynamics in The Zero Eccentricity Limit}
\label{s:ZeroEccentricityLimit}
In the zero \ecc limit, the \xmod phasing formalism reduces
to a dynamical system closely resembling those of the Taylor time-domain
approximants.  The system of ordinary differential equations reduces to 
\beqa
\label{eq:dxdtCircular} 
  M\dot x &=& \frac{64\eta}{5}x^5 \biggl\{ 1 - 
    \left(\frac{743\eta}{336} + \frac{11 \eta^2}{4} \right) x +
      4\pi x^{3/2} \no
          &+& \left(\frac{34\,103\eta}{18\,144} + \frac{13\,661\eta^2}{2\,016} 
                     + \frac{59\eta ^3}{18}\right) x^2 \biggr \}\,, \\
\label{eq:dedtCircular}
  M\dot{e} &=& 0\,, \\
\label{eq:dphidtCircular}
  M \dot{\phi} &=& x^{3/2}\,, \\
\label{eq:dldtCircular} 
  M\dot l &=& x^{3/2} + 3 x^{5/2} + \frac{1}{4}(28\eta-18) x^{7/2} \no
          &+& \left(-7 \eta^2 - \frac{123\pi^2 \eta }{32} + 
              \frac{481\eta }{4} - \frac{27}{2} \right) x^{9/2}\ .
\eeqa
The evolution equations of the system in the zero eccentricity limit are
very similar to the TaylorT4 approximant. The equations which govern the
phase evolution,
(Eqs.~\eqref{eq:dxdtCircular},\eqref{eq:dphidtCircular}) are
equal to the 2~PN TaylorT4 equations~\cite{Buonanno:2009zt}.  The 
eccentric \xmod formalism differs in that it has an 
amplitude contribution entering in through
Eq.~\eqref{eq:dldtCircular}.  We explore the similarity between the two
models in greater depth in Sec.~\ref{s:Results}, where we 
discuss overlap calculations. 

\subsection{Eccentric Binary Waveforms}
\label{s:EccentricWaveforms}
\begin{figure}[ht]
\includegraphics[width=\linewidth]{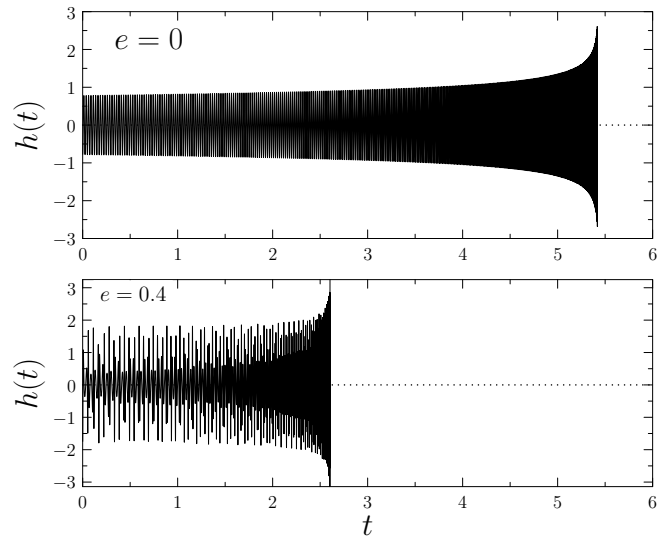} 
\caption{
Gravitational wave signals from a $(1.4+\,10)\,M_{\odot}$ binary 
generated using the \xmod formalism.
The waveforms start at $f_{\mathrm{gw}}=40$ Hz and are
terminated at the Schwarzschild ISCO
frequency. The top panel shows a non-eccentric waveform and
in the bottom panel shows a signal with initial
eccentricity $e_0=0.4$. 
The sharp peaks are due to the increase in
gravitational radiation that
occurs during periastron passage.  The three major effects of
eccentricity illustrated here are: (i) a decreases the duration of
the waveform; (ii) waveform amplitude modulation; and (iii) an overall
increase in amplitude.
} 
\label{fig:waves} 
\end{figure}

We now discuss the eccentric waveforms used for signals
in the template bank simulations. 
The leading order time domain gravitational-wave
polarizations are given by the following expressions:
\bs
\beqa
\label{eq:hplus}
  h_+ &=& \frac{-M\eta}{R}\, \biggl \{ 
    (1+{\cos}^2 \iota) 
      \biggl [ 
        \biggl ( \frac{M}{r} + r^2 \dot{\phi}^2 - \dot r^2 
        \biggr ) 
      \cos2\phi \no 
      &+& 2 r \dot r \dot{\phi} \, \sin 2\phi
      \biggr ] + \,
      \biggl [ 
        \frac{M}{r} -r^2 \dot{\phi}^2 -\dot r^2 
      \biggr ] \sin^2\,\iota
    \biggr \} \,, \\
\label{eq:hcross}
  h_{\times} &=& \frac{-2 M \eta}{R} \cos\iota\, \biggl \{
    \biggl ( 
      \frac{M}{r} + r^2 \dot{\phi}^2 - \dot{r}^2
    \biggr ) 
    \sin2\phi \no
    && - 2 r \dot{r} \dot{\phi} \cos2\phi 
    \biggr \}\,, 
\eeqa
\es
where $R$ is the radial distance of the binary, 
and $\iota$ is the inclination angle of the orbital plane of
the binary measured from the line of sight~\cite{Gopakumar:2001dy,
Thorne:1987af}. When the orbit is circular,
final term in Eq.~\eqref{eq:hplus} vanishes. This can be seen from the
fact that the radial velocity $\dot{r}$ is zero and the Newtonian equation for the
centripetal acceleration is $M/r = r^2 \dot{\phi}^2$.
For circular motion the polarizations simplify to
\beqa
\label{eq:hplusCirc}
  h_+ &=& \frac{-2 M\eta r^2 \Omega^2}{R}\, (1 + {\cos}^2 \iota) \cos2\phi \,, \\
\label{eq:hcrossCirc}
  h_{\times} &=& \frac{-4 M \eta r^2 \Omega^2} {R} \cos\iota\, \sin2\phi \,, 
\eeqa
where $r$ is the binary separation and $\Omega$ is the orbital
frequency.
Notice that we are using
the dominant harmonic only.
Since the post-Newtonian waveforms used here do not capture the merger or
ringdown of the signals, we terminate the waveforms at the Schwarzschild
innermost stable circular orbit (ISCO) frequency, $f_{\mathrm{isco}} = 1 /
\left(6 \sqrt{6} \pi M\right)$.
The waveform $h(t)$ observed at the detector is a linear combination of the
$+$ and $\times$ polarizations:
\beqa
\label{eq:Hoft}
  h(t) &=& F_{+}(\theta,\,\varphi,\,\psi)\, h_{+}(t;\, \iota,\,\phi_0) + \no
    && F_{\times}(\theta,\,\varphi,\,\psi)\, h_{\times}(t;\,\iota,\,\phi_0)\,,
\eeqa
where $F_+$ and $F_{\times}$ are the beam-pattern factors of the
detector and $\phi_0$ is the azimuthal Euler angle of the source.
For ground based interferometers, the beam factors are
expressed in terms of the Euler angles of the detector $(\theta,\,
\varphi)$  and the Euler angle of the polarization plane $\psi$.
Following the convention of Ref.~\cite{Anderson:2001}, the expressions for
the beam-pattern factors are
\bs
\begin{align}
  F_{+}(\theta,\varphi,\psi) &= - \frac{1}{2}\left( 1 + {\cos}^2\theta \right)
       \cos2\varphi \cos2\psi \no
       &-\cos\theta \sin2\varphi \sin 2\psi, \\
  F_{\times} (\theta,\varphi,\psi) &= \frac{1}{2}\left( 1 + {\cos}^2\theta \right) 
      \cos2\varphi \sin2\psi \no
      &- \cos\theta \cos2\varphi\cos2\psi . 
\end{align}
\es
In computing the waveforms, we set the angles such that the binary is
optimally oriented for the $+$ polarization mode; i.e, we set
$F_{+}=1$ and $F_{\times}=0$.   
In Fig.~\ref{fig:waves} we plot
two waveforms of a $(1.4+10) M_{\odot}$ system to illustrate
the effect of \ecc on the waveform.  In comparing the
two waveforms, we find three major effects induced by eccentricity: (i)
amplitude modulation; (ii) decreased
duration of the signal; and (iii) increased signal amplitude.

\section{Data Analysis}
\label{s:DataAnalysis}
LIGO's current searches for gravitational waves from neutron star and stellar-mass black hole binaries~\cite{Abbott:2009qj} use the restricted stationary phase
approximation to the Fourier transform of 3.5~PN circular waveforms (known as
TaylorF2)~\cite{Cutler:1994,Droz:1999qx}. These templates are given in the frequency domain by 
\beq
\label{eq:SPA}
 \tilde{h}(f; M,\eta ) = A(M,\eta)\, f^{-7/6}\,
  \Theta(f-f_c)\, e^{\, i\Psi(f;M,\eta) },
\eeq 
where $A(M,\eta)$ is the overall amplitude of the template at a canonical
distance (typically 1 Mpc), $\Theta$ is the Heaviside step function and $f_c$
is the upper cut-off frequency, given by the Schwarzschild innermost stable
circular orbit (ISCO) frequency, $f_{\mathrm{isco}} = 1 / \left(6 \sqrt{6} \pi
M\right)$. The amplitude is included only at leading order, but the phase
$\Psi(f; M, \eta)$ is computed to 3.5~PN order by
\begin{widetext}
\beqa
  \Psi(f;M,\eta) &=& 2\pi f t_C-2\phi_C-\pi/4 
  + \frac{3}{128 \eta\, v^5} \Biggl \{ 1
  +\left( \frac{3715}{756} + \frac{55}{9} \eta \right) v^2 - 16 \pi v^3 
  + \left( \frac{15\,293\,365}{508\,032} + \frac{27\,145}{504} \eta
  +\frac{3085}{72} \eta^2 \right) v^4 \no
  &+& \pi \left[ \frac{38\,645}{756} -
    \frac{65}{9}\eta \right] \left[1 + 3
    \ln\left( \frac{v}{v_0} \right) \right] 
  + \biggl \{ \frac{11\,583\,231\,236\,531}{4\,694\,215\,680} -
\frac{640}{3} \pi^2 
  -\frac{6\,848}{21} \left(\gamma+\ln(4\,v)\right) \no
  &+& \left(-\frac{15\,335\,597\,827}{3\,048\,192} +
    \frac{2\,255}{12} \pi^2 \right) \eta 
  + \frac{76\,055}{1\,728} \eta^2
  - \frac{127\,825}{1\,296} \eta^3 \biggr \} v^6 \no
  &+& \pi \left[ \frac{77\,096\,675}{254\,016} + \frac{378\,515}{1\,512}
    \eta - \frac{74\,045}{756} \eta^2 \right] v^7 \Biggr \}\,,
\eeqa
\end{widetext}
where $v = \left(\pi M f \right)^{1/3}$ and
$\gamma$ is Euler's constant~\cite{Blanchet:2001ax,Blanchet:2004ek}.
To search for a signal in detector data, we construct the
signal-to-noise ratio $\rho$ from the template $h$ and the calibrated
output of the detector $s$ according to
\begin{equation}
  \rho = \frac{1}{\sqrt{\bra h \mid h \ket}} \bra s \mid h \ket,
\end{equation}
where the inner product is given by 
\beqa
\label{eq:InnerProduct}
  \bra s \mid h \ket & = & 2\, \int_{0}^{\infty} \ud f\,
    \frac{ {\tilde{s}}^*(f) \tilde{h}(f) +
      \tilde{s}(f) {\tilde{h}}^*(f) }
           { S_n (f) }, \no
        & = & 4 \, \Re \int_{0}^{\infty} \ud f\, 
    \frac{ {\tilde{s}}^*(f) \tilde{h}(f) }
     { S_n (f) }\,,  
\eeqa
$\tilde{s}(f)$ denotes the Fourier transform of $s(t)$, 
\beq
\label{eq:FourierTrans}
  \tilde{s}(f) = \int_{-\infty}^{\infty} \ud t\, e^{-2\pi i f t} s(t),
\eeq
and $S_n(f)$ is the one-side noise power spectral density (PSD) of the
detector. In the presence of Gaussian noise alone, the expectation value of
$\rho^2$ is unity; large values of $\rho$ indicate that a signal is present in
the data. The template parameter $t_C$ gives the time of arrival of the
signal, and can be searched over using a Fourier
transform~\cite{Owen:1995tm}. Similarly the coalescence phase $\phi_C$ can be
searched over analytically~\cite{Allen:2005fk}; these parameters are
termed extrinsic parameters.
The masses of the signal are not known \emph{a priori}, however, and so a
discrete bank of templates must be constructed~\cite{Owen:1998dk} and the
detector data filtered against each template in the bank. Since the mass
parameters $M, \eta$ must be explicitly searched over, they are known as intrinsic parameters.
The \emph{ambiguity function} $\mathcal{A}(\vec{\theta})$ is defined with respect to the template
parameters $\vec{\theta} = ( t_C,\phi_C,M,\eta )$ as 
\beq
\label{eq:Ambiguity}
  \mathcal{A}(\vec{\theta}) = \frac{ \bra s \mid h \ket }{ \sqrt{
\bra s \mid s \ket \bra h \mid h \ket } }.
\eeq
The maximization of Eq.~\eqref{eq:Ambiguity}
over the extrinsic template parameters defines the overlap
\beqa
\label{eq:Overlap}
   \mathcal{O}(s,h) & = & \max_{ t_C,\,\phi_C } \,
     \frac{ \bra s \mid h \ket }{ \sqrt{ \bra s \mid s \ket \bra h \mid h \ket } }, \no 
     & = & \max_{ t_C,\,\phi_C }\, \bra \vs \mid \vh \ket,
\eeqa
\cite{Owen:1995tm, Owen:1998dk}, where the hat denotes normalization of the
template amplitude.
The fitting factor $\FF$,
quantifies loss in signal-to-noise ratio (SNR) due to the
filtering of a signal with template that is not an optimal filter~\cite{Apostolatos:1994mx}.
The loss in event rate incurred by
non-optimal filtering is proportional to $1-{\FF}^3$. For instance, a 3\%
decrease in $\FF$ gives a loss of approximately 10\% in event rate.
The value of $\FF$ is obtained by
maximizing the overlap function over the intrinsic parameters of the
template waveform 
\beq
\label{eq:FitFact}
  \mathcal{F} = \max_{M, \, \eta} \max_{t_C, \, \phi_C}  
  \bra \hat{s} \mid \hat{h} \ket\,.
\eeq
When constructing a bank, the templates are typically placed such that the
loss in signal-to-noise ratio due to mismatch between a template with
parameters inside the bank and the nearest bank grid point is no greater than
3\%; such a bank has a fitting factor of 0.97.

In this paper, we use the
hexagonal template placement algorithm from the LIGO Algorithm Library (LAL)~\cite{LAL,Babak:2006ty}
to place a bank which runs over the mass range $ 1 \leq m_1,\,m_2 \leq 34 M_{\odot} $ subject to $ M \leq
35 M_{\odot}$ at a fitting factor of 0.97. 
Template placement depends on the shape of the
noise power spectrum of the
detector~\cite{Owen:1998dk}. We model the current LIGO detectors by an analytic fit to the LIGO
noise curve~\cite{InitialLIGONoise}, given by
\beqa
  S_n(f) &= & 9.0 \times 10^{-46} 
  \bigg [
    \left( 4.49 x \right)^{-56} \nonumber \\
    &+ & 0.16 x^{-4.52} + 0.32 x^2 + 0.52
  \bigg ]\, \mathrm{Hz}^{-1}\,,
\eeqa
where $x = f/150\mathrm{Hz}$ and $f$ is frequency in Hz. Since the current LIGO
detectors (Enhanced LIGO) have the same seismic isolation as the Initial LIGO detectors,
they have essentially the same low frequency response.
Therefore the Initial LIGO PSD is sufficient to model the Enhanced LIGO
detectors for the purpose of this study.  The AdvLIGO
PSD~\cite{AdvancedLIGONoise} is modeled by
\begin{align}          
\label{eq:AdvLIGOPsd}
S_n(f) &= 1.60\times10^{-49}\, \no
       & \biggl[ 300 \left(\frac{f}{15\mathrm{Hz}}\right)^{-17} 
            + 7 \left(\frac{f}{50\mathrm{Hz}}\right)^{-6} \no
       &+ 24\, \left(\frac{300\,x}{90}\right)^{-3.45} - \frac{3.5}{x^2} \no
       &+ 110\, \left(\frac{1.02-1.08\,x^2+0.54\,x^4}{1+0.21\,x^2}\right) 
         \biggr]\, \mathrm{Hz}^{-1}\,,
\end{align}
where $x=f/300\mathrm{Hz}$.
The low-frequency cutoff $f_0$ for the templates is 40 Hz for LIGO and 10 Hz
for AdvLIGO. 

To study the effectiveness of the template bank described above to capture
signals from eccentric systems, we model eccentric binaries by the waveform
given in Eq.~\eqref{eq:Hoft}, which includes 3~PN conservative dynamics
and 2~PN
radiative dynamics. 
The binary is chosen to be optimally oriented~\cite{Thorne:1987af}.
We generate an eccentric signal and compute the overlap of the waveform
against each template in the bank. The overlap is maximized over the template
bank to obtain the best match. Following Ref.~\cite{Lindblom:2008cm}, we
define the \emph{effective fitting factor} $\bar{\mathcal{F}}$
\beq
\label{eq:EffFitFact}
  \bar {\mathcal{F}} = \max_{\hat{h}\,\in\,\mathrm{bank}} \bra
                         \hat{s} \, \mid \, \hat{h} \ket .
\eeq
The function $\bar{\mathcal{F}}$ provides a measure of how effective the
template bank is in capturing a signal $s$ that does not lie exactly in the
template bank manifold.  The distribution of $\bar {\mathcal{F}}$  allows one
to isolate the regions of the physical parameter space where the
mismatch is less than the minimal match of the bank.  

\section{Results}
\label{s:Results}

\subsection{Initial and Enhanced LIGO}
\label{s:iLIGOResults}
\begin{figure}[htpb]
\includegraphics[width=0.95\linewidth]{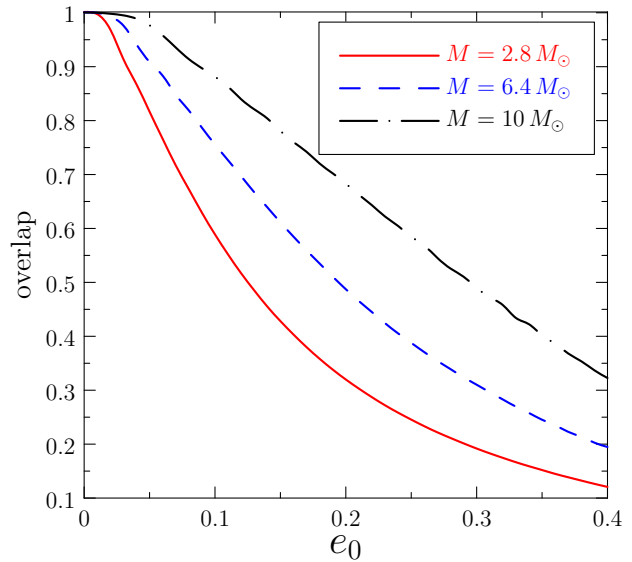}
\caption { 
The overlap $\bra \hat{s}_{e\,=\,0}\,|\hat{s} \ket$ using the Initial
LIGO PSD for three systems having total masses
$ (1.4 + 1.4) M_{\odot},\, (1.4 + 5.0) M_{\odot} $ and $(5.0+5.0)
M_{\odot} $.
Here, $s_{ e\,=\,0 }$ denotes the circular \xmod
signal. The overlaps for the higher mass system are less affected by
eccentricity because they have fewer cycles in the LIGO frequency
band, and can thus be phase-matched against template
waveforms via a relative shift in coalescence time without significant
phase decoherence.
}
\label{fig:SELFolaps} 
\end{figure}

\begin{figure}[htpb]
\includegraphics[width=0.95\linewidth]{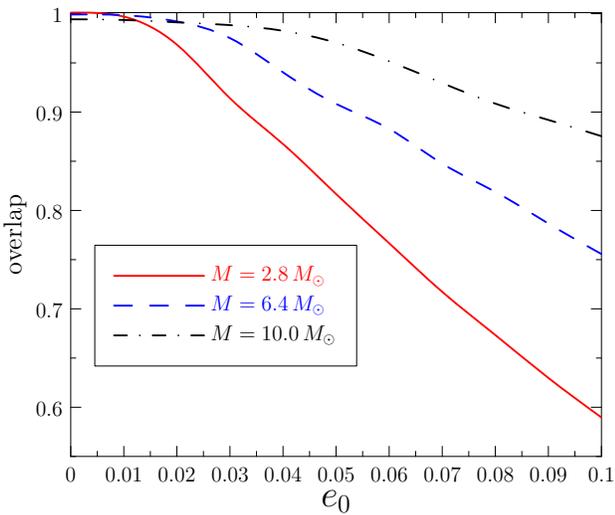}
\caption{ 
Overlaps using the LIGO PSD between the 2~PN \xmod waveform and the 2~PN TaylorT4 waveform of
the same total mass for masses $ (1.4+1.4)\,M_{\odot},\, (1.4+5.0)\,M_{\odot} $ 
and $(5.0+5.0)\,M_{\odot} $. 
The overlaps approach one as the eccentricity goes to zero, which
indicates that the two models are in good agree in the limit $e_0 \rightarrow 0$. 
}
\label{fig:T4olaps} 
\end{figure}

In this section we present the results of our
study for the current LIGO detectors.
To understand the degree to which eccentricity affects the
the \xmod waveform, we investigated the
overlap as a function of eccentricity using a circular ($e_0=0$) \xmod waveform
as the template and a \xmod waveform with non-zero eccentricity as the signal.  
Fig.~\ref{fig:SELFolaps} shows the LIGO PSD overlap
as a function of initial eccentricity for three different mass ratios. 
The overlaps are more sensitive to eccentricity for smaller mass
systems, since the lower mass systems have longer durations which allow for the
effects of eccentricity to accumulate. 
We next investigate overlaps between \xmod signals and 2~PN TaylorT4
templates.  In section ~\ref{s:ZeroEccentricityLimit}, we noted the
similarities between the zero eccentricity limit of the $x$-model and
the 2~PN TaylorT4 approximation. Comparisons are made with the TaylorT4
at 2~PN since that is the radiative PN order at which the $x$-model has been computed.  In Fig.~\ref{fig:T4olaps},
we plot the LIGO PSD overlap between 2PN TaylorT4 waveforms and \xmod waveforms
of the same mass as a function of eccentricity.  As the eccentricity approaches zero, the
overlaps between the T4 and the \xmod approach unity for all three
signal masses, consistent with the zero eccentricity
limit of the $x$-model in Sec.~\ref{s:ZeroEccentricityLimit}.

\begin{figure}[htpb]
\includegraphics[width=0.95\linewidth]{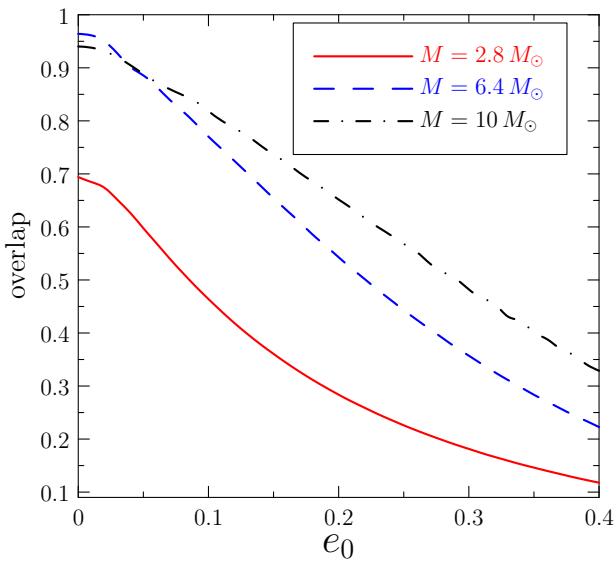} 
\caption{
Initial LIGO overlaps between the 2~PN \xmod and the 3.5~PN TaylorF2 waveforms as a function of
initial eccentricity for $(1.4+1.4)\,M_{\odot},\,(1.4+5.0)\,M_{\odot}$
and $(5.0+5.0)\,M_{\odot}$ systems.  Notice that the $e_0=0$ overlaps
are not unity.  This is due to differences in the PN order of the
waveforms
}
\label{fig:olap3p5spa}
\end{figure}

\begin{figure}[t]
\includegraphics[width=0.95\linewidth]{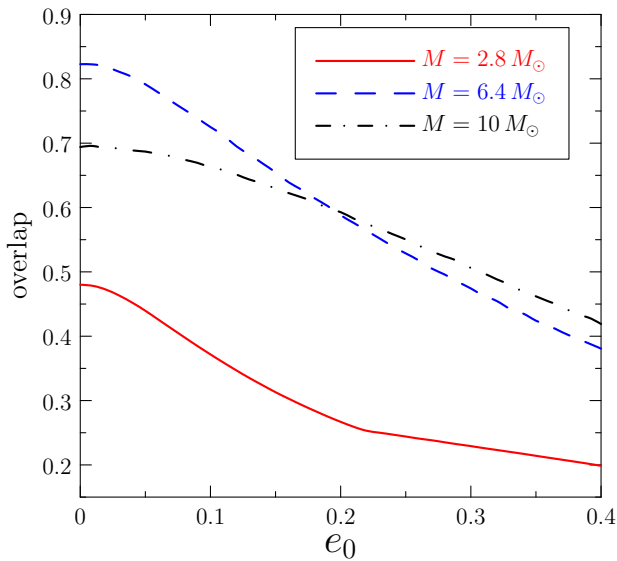} 
\caption{
Overlaps between the 2~PN $x$-model eccentric waveforms and the
3.5~PN TaylorF2 waveforms using the AdvLIGO PSD.
Reduced overlaps result from the increase in the amount
of time spent in the detector's band. The increase in the duration
leads to larger phase de-coherence and intensifies the effects due to
differences in PN order.  Additionally, the difference in the PN order
of the signals and the templates reduces the overlap even at $e_0=0$.  
} 
\label{fig:olapsAdvanced}
\end{figure}

Fig.~\ref{fig:olap3p5spa} shows overlaps for Initial LIGO between 2~PN
\xmod waveforms and 3.5~PN TaylorF2 templates. The overlap is worse for
these waveforms, as would be expected. At $e_0 = 0$ the reduction in overlap
is due to the difference between the PN models used to
generate waveforms used as signals and
templates. As $e_0$ increases, the figure shows the additional loss in
overlap due to the effect of eccentricity. The fact that the overlaps are not unity at $e_0 =
0$ is due to the differences between the PN formulation of the time-domain
$x$-model signal, which contains radiative effects to 2~PN order, and the
frequency-domain TaylorF2 template, which contains radiative effects to 3.5~PN
order. Although higher-order post-Newtonian effects are more significant (at a
given frequency) for higher-mass binaries, more massive binaries coalesce at
lower frequencies. At $e_0 = 0$, the lowest overlap is observed in $M = 2.8\
M_\odot$ signal. Although higher order post-Newtonian corrections are small for
this system, they have many waveform cycles over which to accumulate,
resulting in a reduction of overlap.  Note, however, that in calculating the
overlaps shown in Figs. \ref{fig:olap3p5spa} and
\ref{fig:olapsAdvanced} the intrinsic parameters are kept fixed. Below we show the effect of
maximizing the overlap over a bank, as is the case in a real search. As $e_0$
increases, eccentricity causes additional higher-order post-Newtonian
corrections to the signal waveform, but also causes the duration of the
waveform to decrease. In the $M = 6.8\, M_\odot$ and $M = 10\, M_\odot$
waveforms, competition between these effects causes the overlap curves to
cross at $e_0 = 0.05$, with slightly larger overlaps in the higher-mass signal
at larger eccentricity.

\begin{figure*}[thb]
\includegraphics[width=0.32\linewidth]{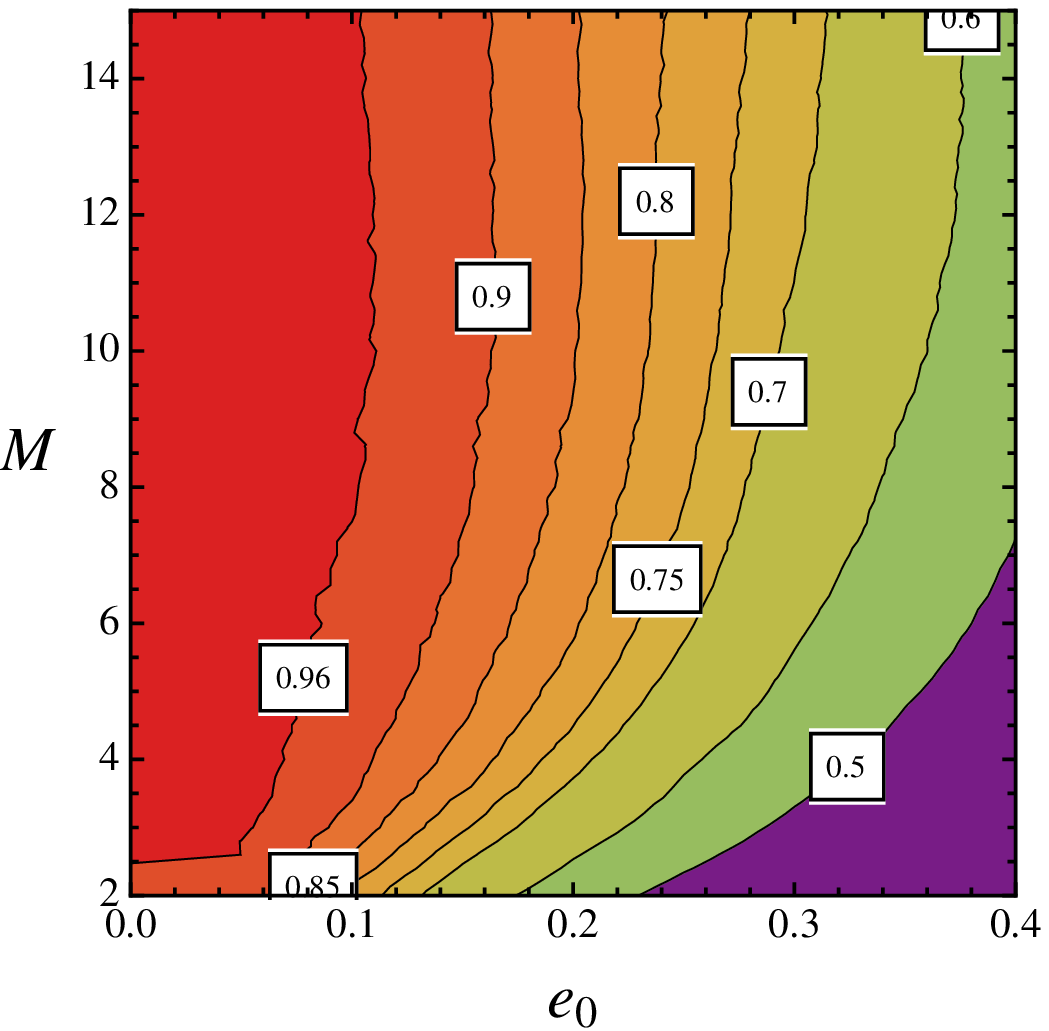} 
\includegraphics[width=0.32\linewidth]{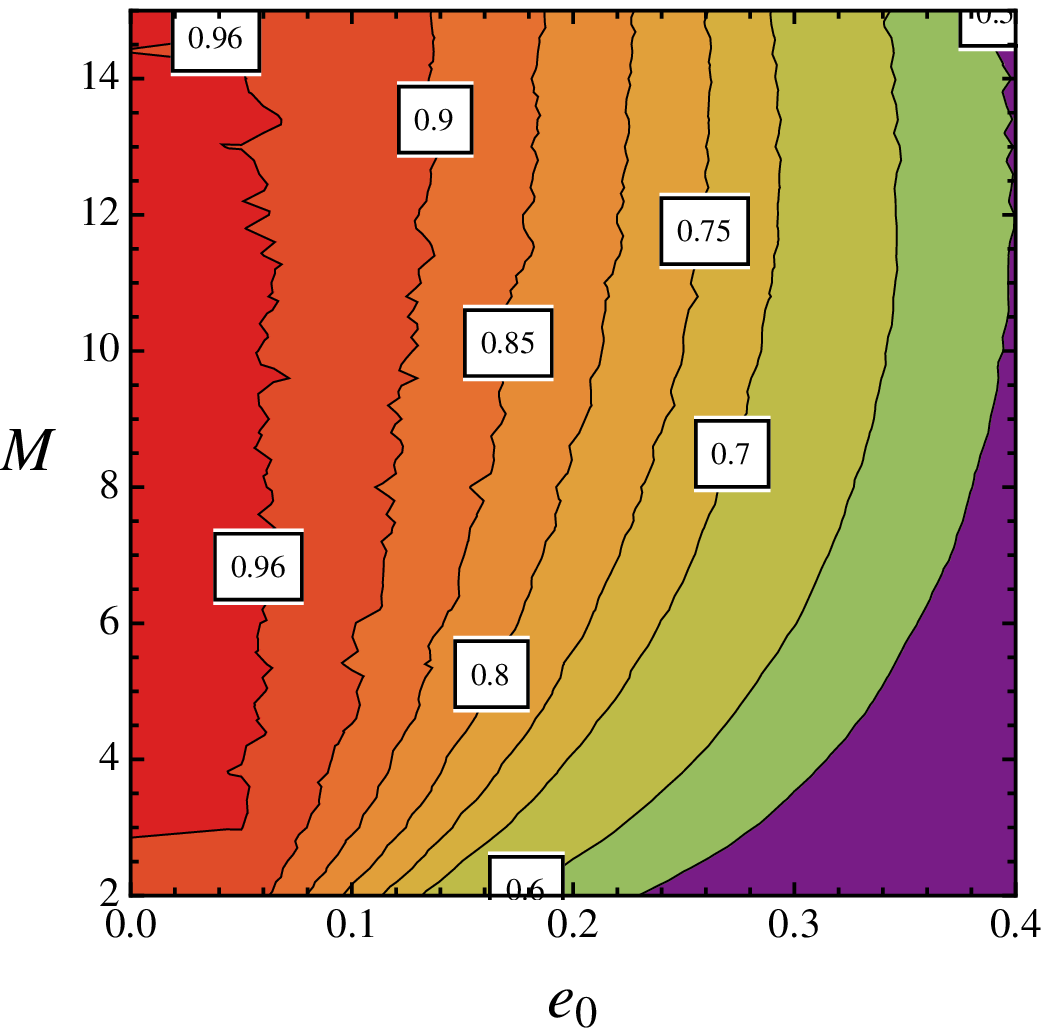} 
\includegraphics[width=0.32\linewidth]{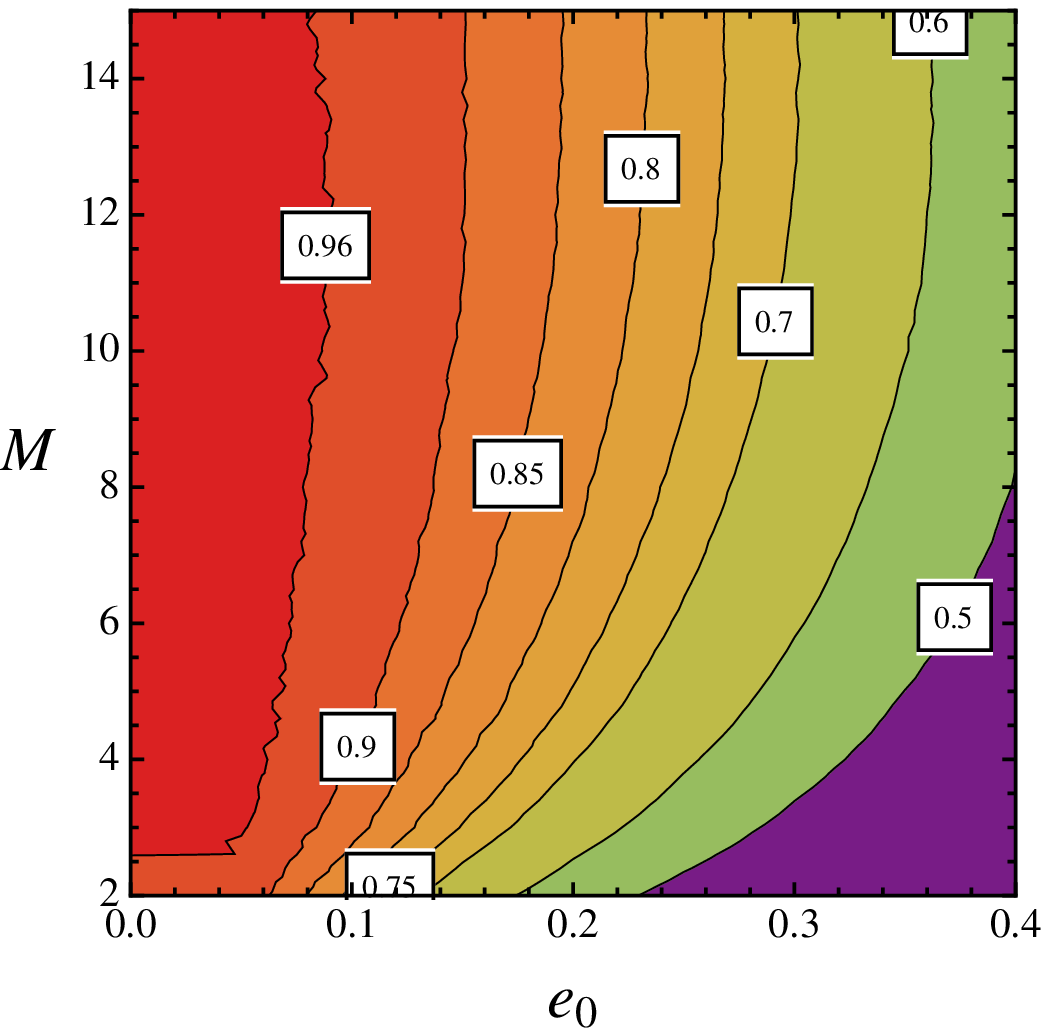} 
\caption{
From left to right the figures show the best, worst and mean value of the
effective fitting factor $\bar\FF$ for the initial LIGO 3.5 PN TaylorF2 bank
as a function of the total mass $M$ in solar masses and the eccentricity
$e_0$ at $40$~Hz of the injected signal. The contours are labeled by the values
$\bar\FF$ enclosed.  
}
\label{fig:iLIGOmatch}
\end{figure*}

\begin{figure*}[htb]
\includegraphics[width=0.32\linewidth]{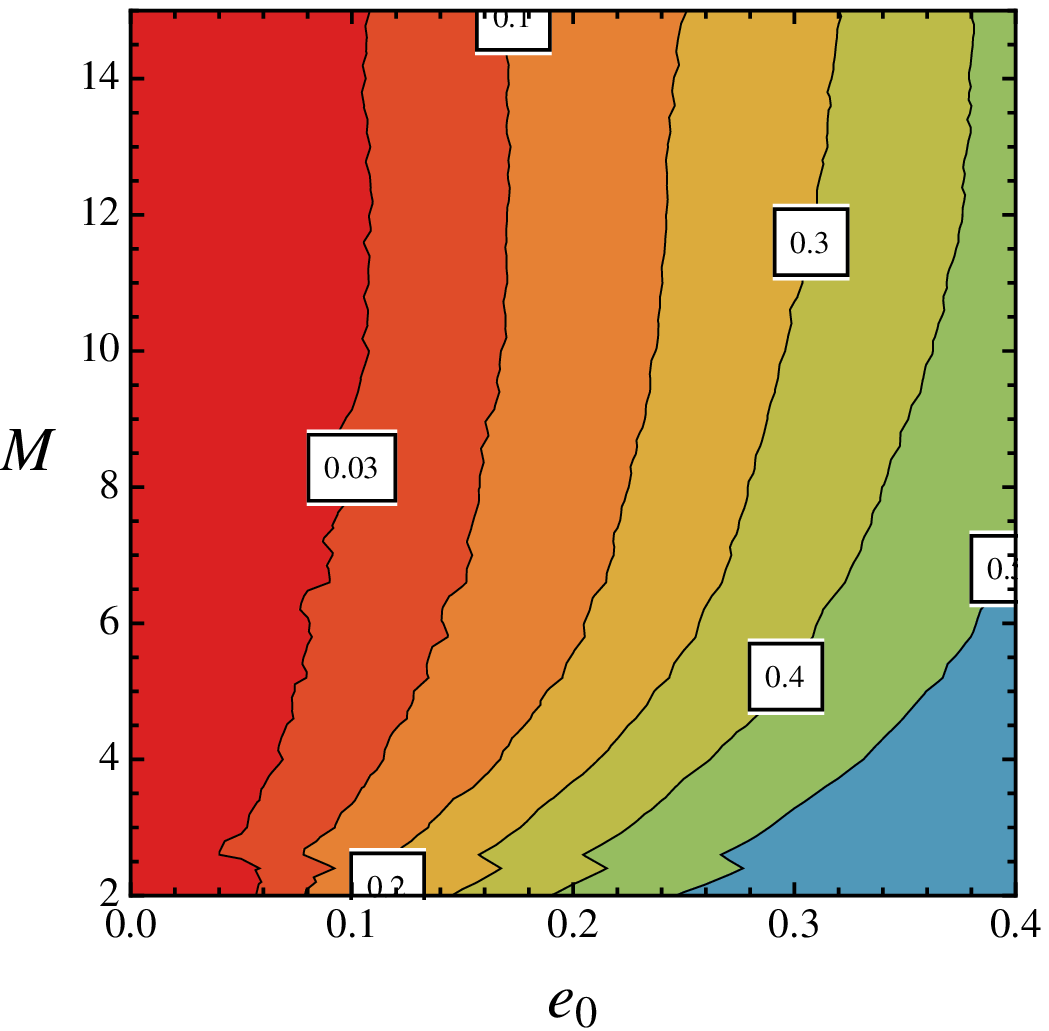}  
\includegraphics[width=0.32\linewidth]{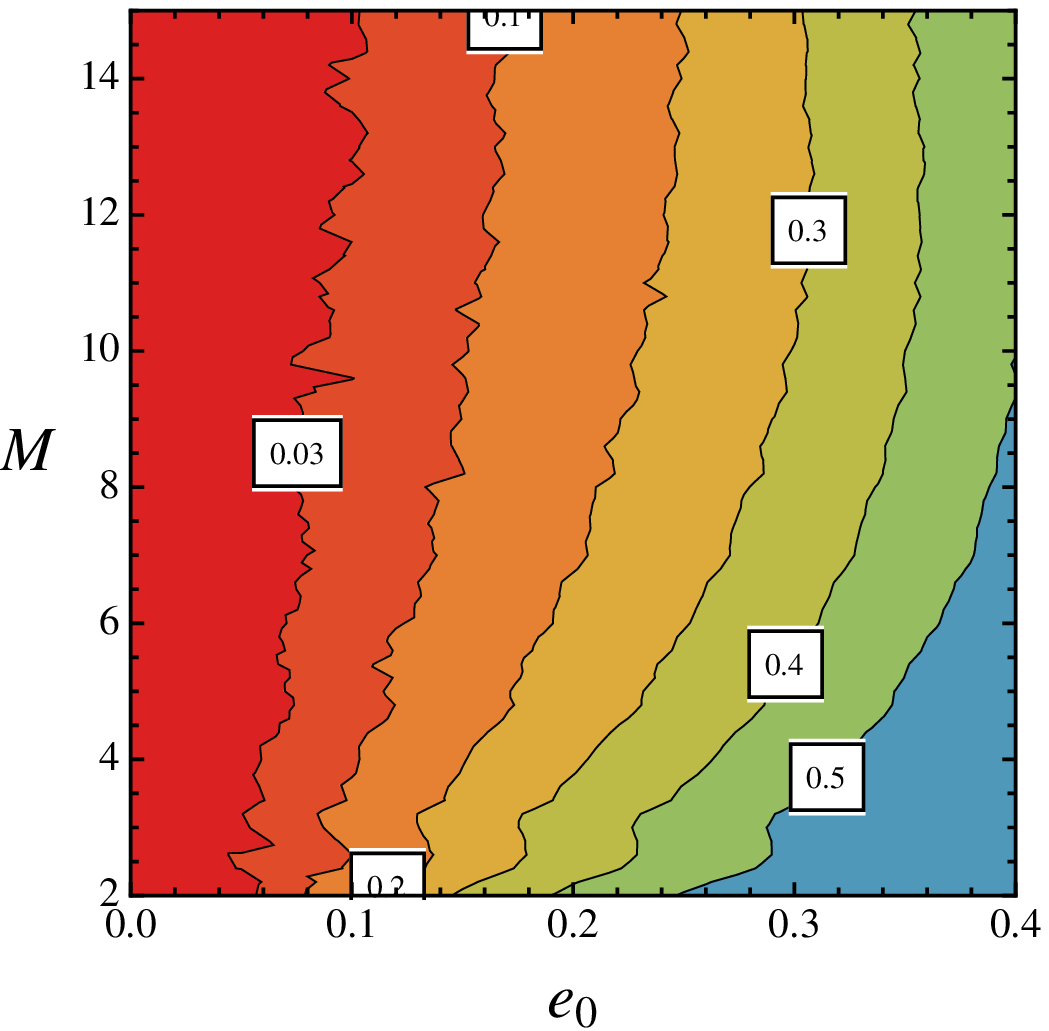} 
\includegraphics[width=0.32\linewidth]{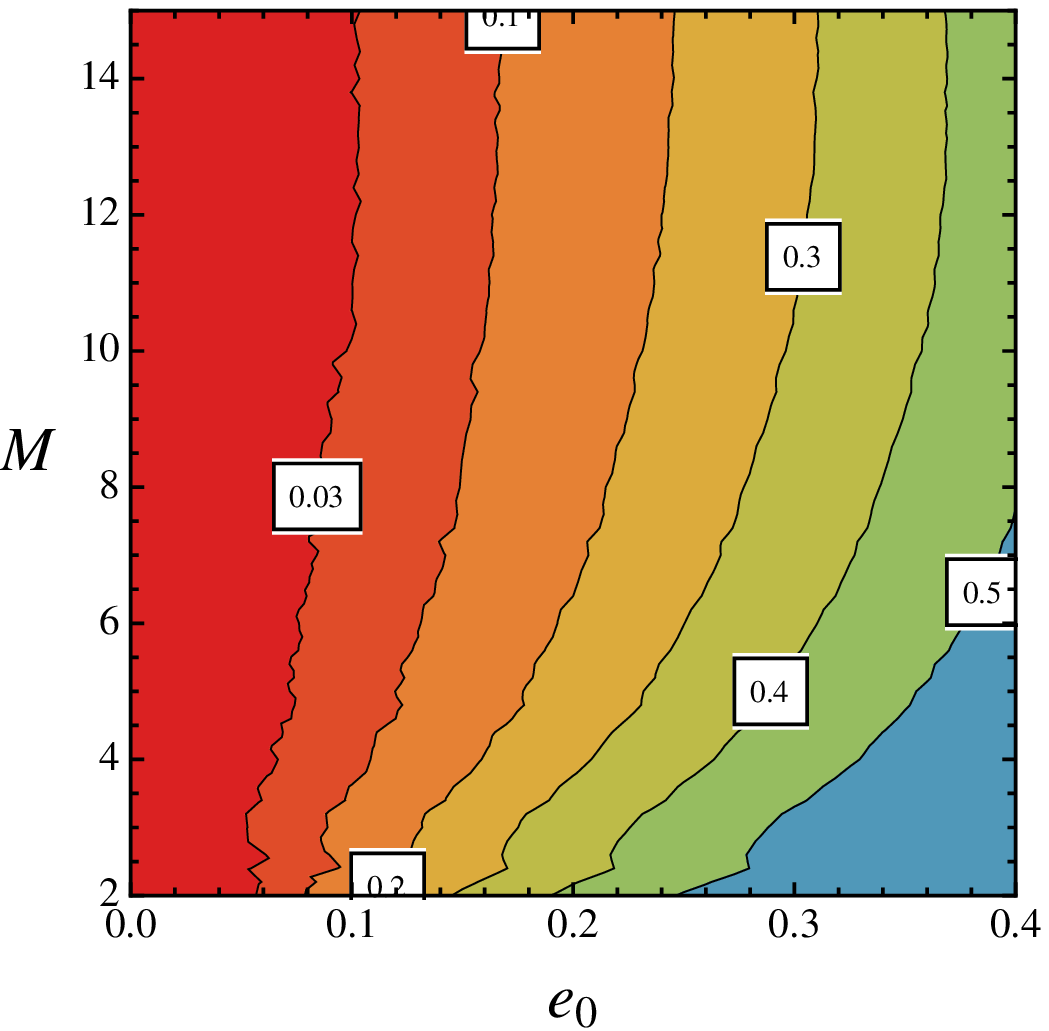} 
\caption{
From left to right the figures show the best, worst and mean value of
$\Delta\bar\FF$ for the Initial LIGO 3.5 PN TaylorF2 bank as a function of the
total mass $M$ in solar masses and the eccentricity $e_0$ at $40$~Hz of the
injected. The contours are labeled by the values $\Delta\bar\FF$ enclosed.
}
\label{fig:iLIGOloss}
\end{figure*}

We now explore the effectiveness of the template bank described in
Sec.~\ref{s:DataAnalysis} at detecting eccentric signals. To explore this for
a range of signal masses and eccentricities, we generate a rectangular grid of
$10\,026$ $x$-model signals with masses in the range 
$ 1\Msun \leq m_1,\,m_2 \leq 14\Msun $ subject to $ 2 \leq M/M_{\odot} \leq 15 $ and initial eccentricities in the range $ 0 \leq e_0 \leq 0.4 $. Each
signal in this grid is filtered through the $14\,863$ templates in the 3.5~PN TaylorF2
Initial LIGO template bank. The effective fitting factor $\bar\FF$ of the bank
is computed for each signal. The signal space is three dimensional as it
depends on $m_1,m_2,$ and $e_0$, and hence $\bar\FF$ is a function of three variables.
To visualize the results of the bank simulation, we plot $\bar\FF$ as a
function of the total mass $M$ and eccentricity $e_0$ of the injected $x$-model
signal.  In our signal grid, several component masses have the same total mass
and so in Fig.~\ref{fig:iLIGOmatch}, we plot the highest, lowest, and mean
$\bar \FF$ for each value of $M$.
Notice that by maximizing over the template banks, we have gained
significantly in overlap for the lower mass systems (cf.
Fig.~\ref{fig:olap3p5spa}). In the worst case, the 3.5 PN TaylorF2 bank can
achieve effective fitting factors $\ge 0.96$ for all $x$-model signals which
have a total mass $M\ge 3\Msun$ and an eccentricity $e_0 \le 0.05$ at 40~Hz. 
The signals with the worst values of
$\bar\FF$ are those which are matched against templates which lie near the edges of the template bank. 
To separate the effect of eccentricity from effects related to the difference
in PN order between the signal and template we calculate the relative
difference between the effective fitting factor for signals with zero
eccentricity and those with non-zero eccentricity, given by
\begin{equation}
  \Delta\bar\FF(e_0) = \frac{\bar\FF(e_0=0)-\bar\FF(e_0)}{\bar\FF(e_0=0)}.
\end{equation}
Fig.~\ref{fig:iLIGOloss} shows $\Delta\bar\FF$ for the highest, lowest, and
average values of $\bar\FF$. These distributions illustrate that the loss in
overlap results predominantly from eccentricity at higher masses, but at lower
masses is the accumulation of higher-order PN effects (included in the
templates, but not in the signals) over the longer waveforms.
We therefore conclude that the current LIGO template bank is effective at
capturing signals with small residual eccentricities $e_0 \lesssim 0.05$ at
$40$~Hz for all masses $2 \le M \le 15\Msun$. For eccentricities higher
than $e_0
\sim 0.05$, the loss in $\bar\FF$ becomes more significant, although for the
higher mass systems this loss does not occur until $e_0 \gtrsim 0.1$. 

\subsection{Results for Advanced LIGO}
\label{s:advLIGOResults}
Finally, we investigate how eccentricity effects the detection efficiency of
TaylorF2 templates for AdvLIGO. Although it is unlikely that TaylorF2
templates will be used for AdvLIGO searches, this study is illustrative of
the effect of eccentricity on circular templates with the AdvLIGO noise
curve.  The lower frequency cut-off for the AdvLIGO PSD requires that we start
our waveforms at 10 Hz instead of 40 Hz.  
Fig.~\ref{fig:olapsAdvanced} shows the overlap between the 3.5 PN TaylorF2
waveforms and the eccentric $x$-model waveforms for the AdvLIGO noise curve.
Since the waveforms have more detectable cycles in AdvLIGO, a larger phase
difference accumulates leading to lower overlaps than for Initial
LIGO.  The effect of the post-Newtonian differences between the waveform and
templates is also apparent in Fig.~\ref{fig:olapsAdvanced}, as in
Fig.~\ref{fig:olap3p5spa}.

\begin{figure*}[thb]
\includegraphics[width=0.32\linewidth]{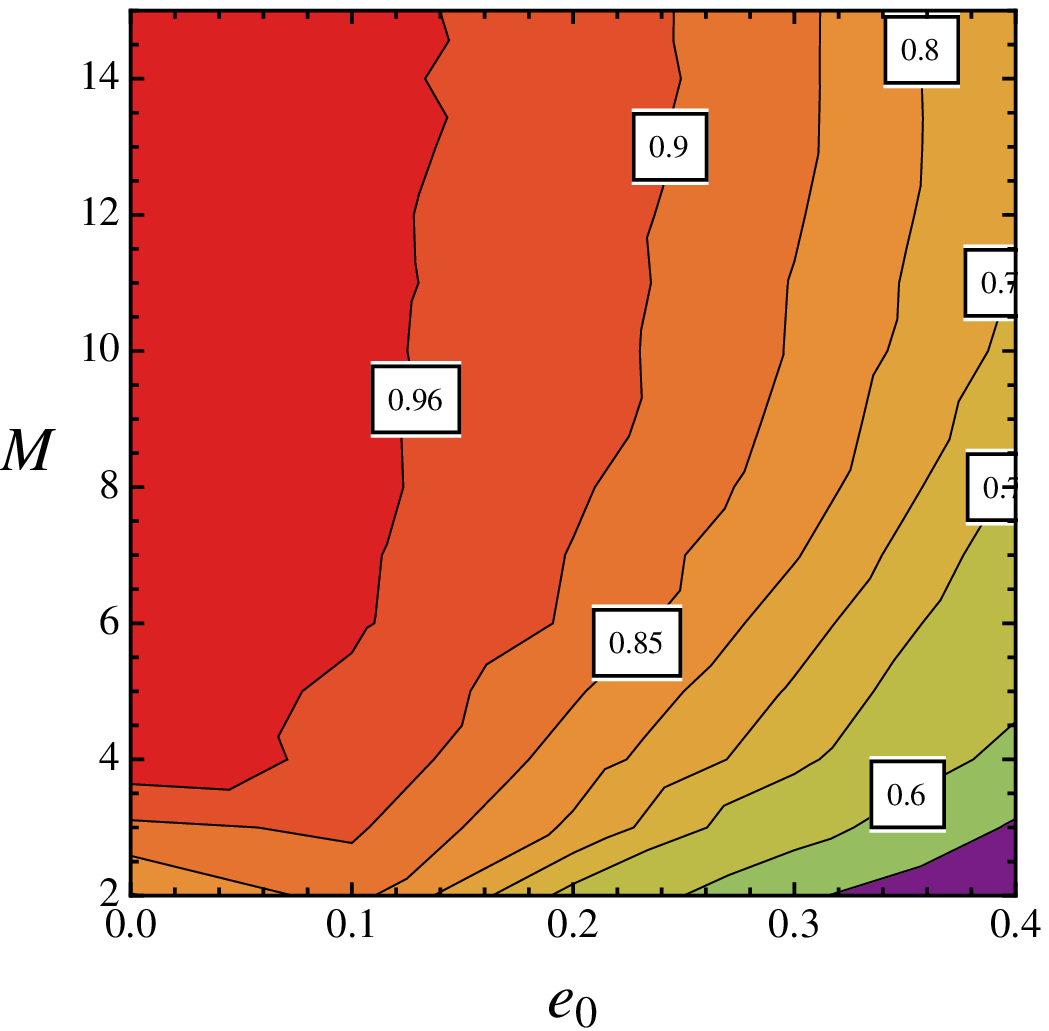} 
\includegraphics[width=0.32\linewidth]{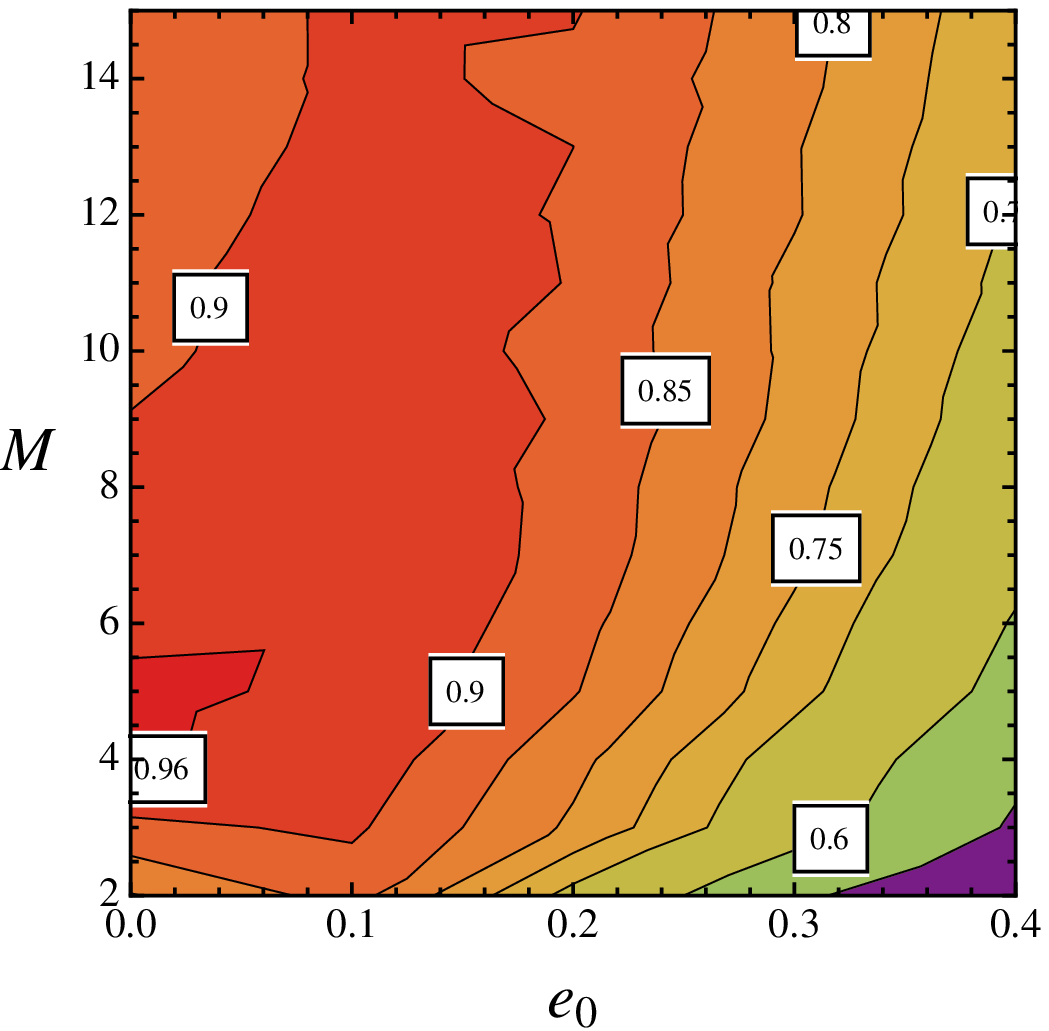} 
\includegraphics[width=0.32\linewidth]{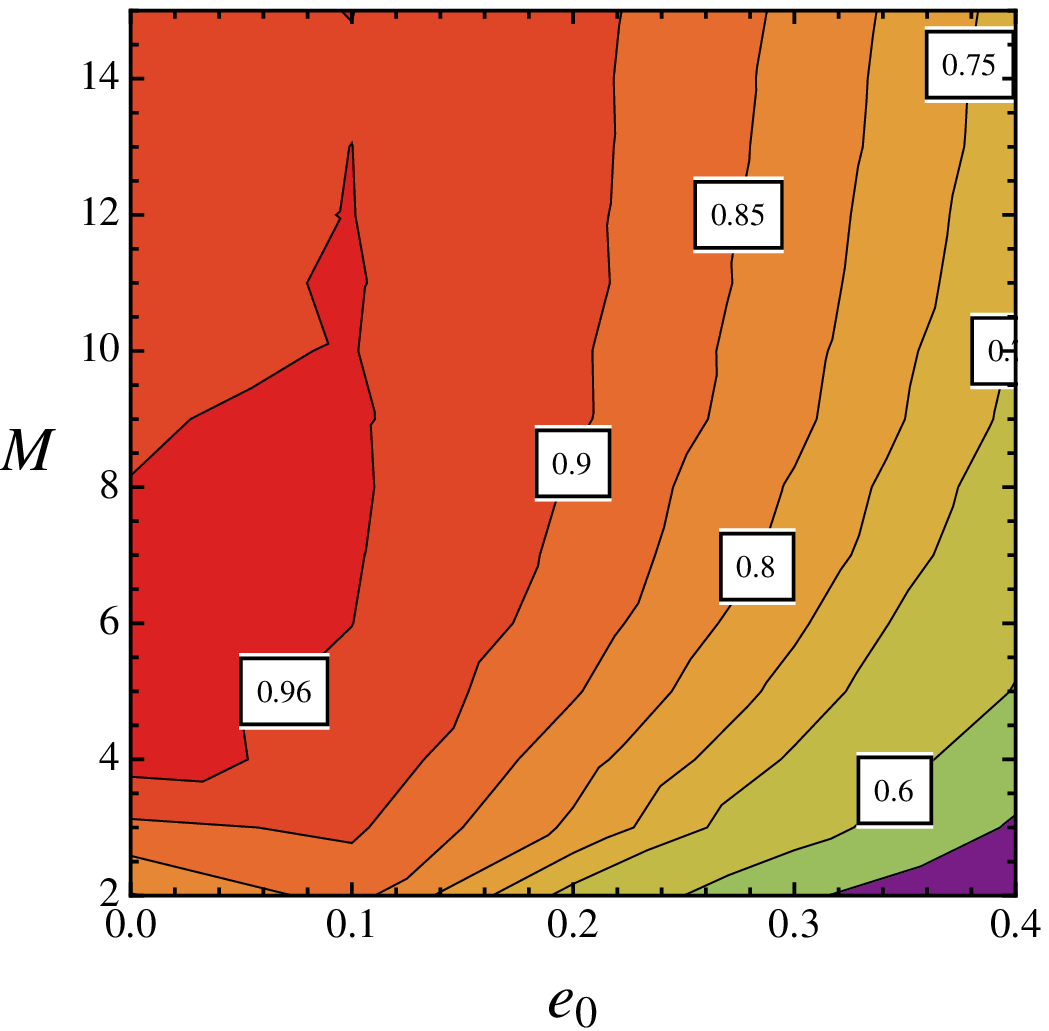} 
\caption{
From left to right the figures show the best, worst and mean value of the
effective fitting factor $\bar\FF$ for the AdvLIGO 3.5~PN TaylorF2 bank
as a function of the total mass $M$ in solar masses and the eccentricity
$e_0$
at $40$~Hz of the injected signal. The contours are labeled by the values
$\bar\FF$ enclosed.  
}
\label{fig:advLIGOmatch}
\end{figure*}

\begin{figure*}[htb]
\includegraphics[width=0.32\linewidth]{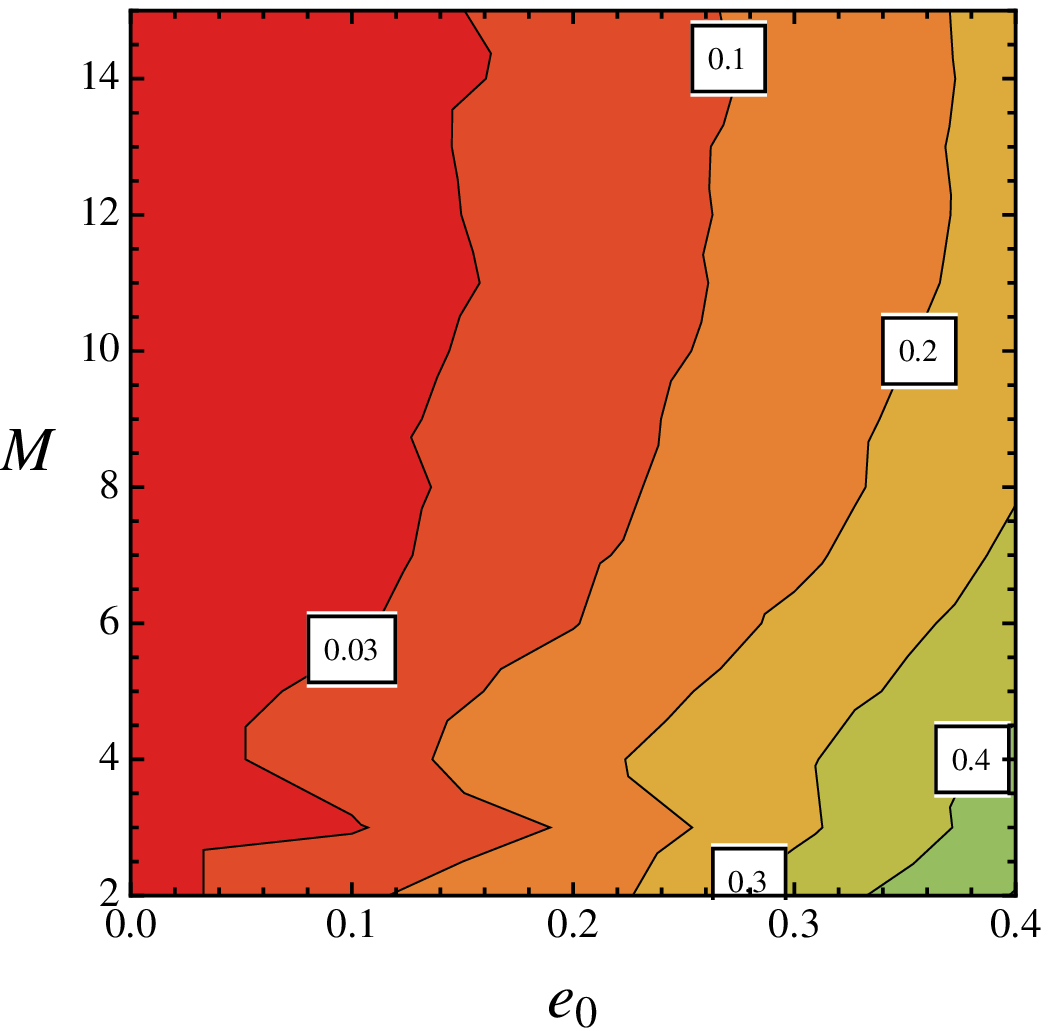} 
\includegraphics[width=0.32\linewidth]{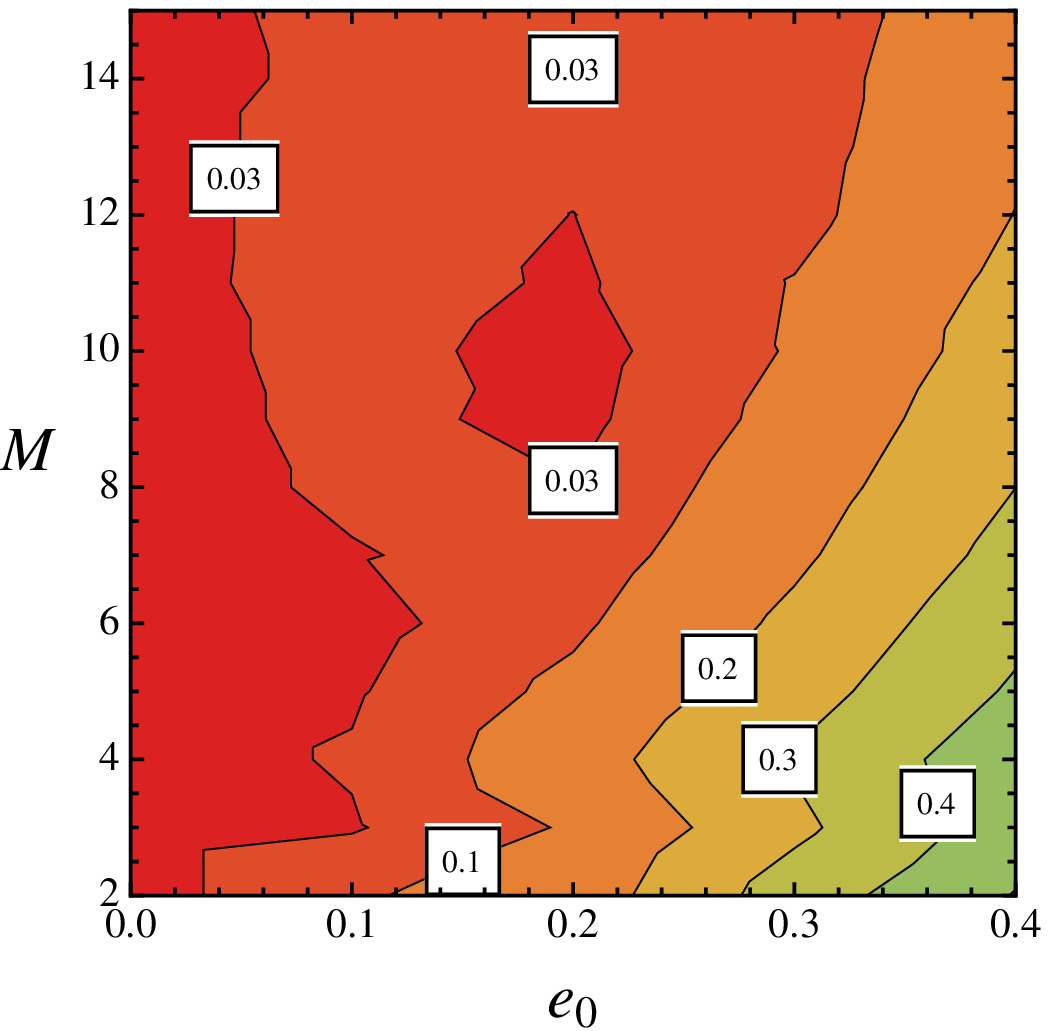} 
\includegraphics[width=0.32\linewidth]{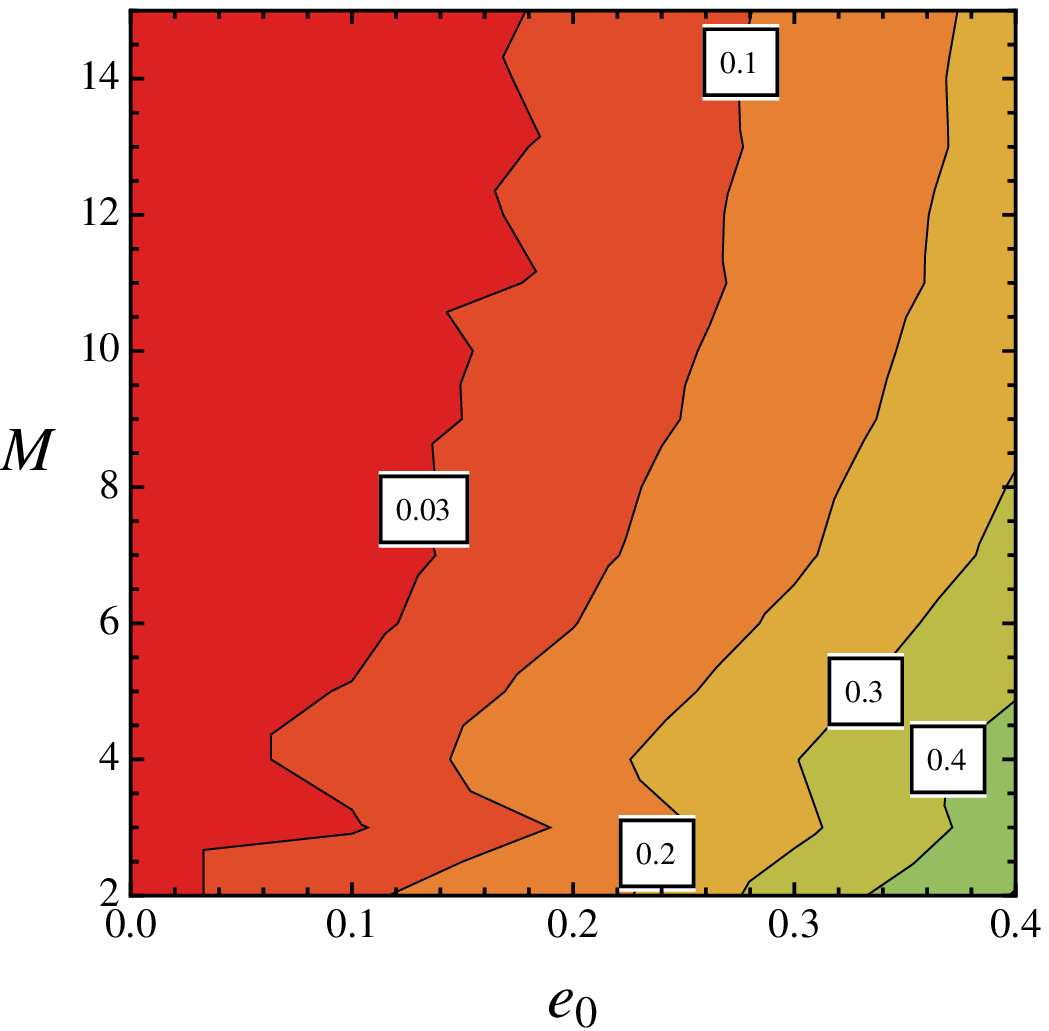} 
\caption{
From left to right the figures show the best, worst and mean value of
$\Delta\bar\FF$ for the AdvLIGO 3.5~PN TaylorF2 bank as a function of the
total mass $M$ in solar masses and the eccentricity $e_0$ at $40$~Hz of the
injected signal. The contours are labeled by the values $\Delta\bar\FF$ enclosed.
}
\label{fig:advLIGOloss}
\end{figure*}

Our analysis of the AdvLIGO template bank results proceeds in a similar way to
that for the current detectors. Due to the improved low-frequency sensitivity
of AdvLIGO, the size of the template bank increases to $136\,000$, greatly
increasing the computational cost of the template bank simulation. As a
result, we decrease the resolution of the signal grid to inject only 280
signals. This is sufficient to probe the effect of eccentricity on AdvLIGO
searches, however.  Fig.~\ref{fig:advLIGOmatch} shows the best, worst and mean
values of $\bar\FF$. 
Notice that $\bar\FF$ is very poor for the long, low-mass signals,
and the high-mass signals (where higher order PN effects
dominate) even at zero eccentricity.  This is due to the difference
in PN order between the signals and the templates.
In Fig.~\ref{fig:advLIGOloss} we plot $\Delta\bar\FF$ for the highest, lowest,
and average values of $\bar\FF$ for AdvLIGO. In the AdvLIGO case, 
 the effect of small residual eccentricity can be significant for low
mass systems $(M\lesssim 4\Msun)$.
Further studies using both higher PN-order eccentric waveforms and
the actual AdvLIGO search templates are needed to confirm these
findings.

\section{Conclusion}
\label{s:Conclusion}
In this paper we studied the effect of eccentricity on detection searches for
gravitational waves from compact binary coalescence using the current and
Advanced LIGO detectors.  To construct eccentric binary waveforms, we used the
NR-calibrated 2~PN \xmod formalism proposed
by Hinder \emph{et al.}~\cite{Hinder:2008kv}. 
We considered the zero eccentricity
limit of the $x$-model and compared it with the TaylorT4 approximant;
analytic and overlap calculations confirm that the
phase evolutions of these two models agree in the $e_0=0$ limit.    
Using \xmod waveforms as our signal family, we performed template bank simulations
using the 3.5 PN TaylorF2 waveforms and the hexagonal placement algorithm implemented in
LAL~\cite{LAL}.  The 3.5 PN TaylorF2
template bank was found to give fitting factors of 0.96 or better for systems
with total mass $M \gtrsim 3M_\odot$ and eccentricity at $40$~Hz of $e_0
\lesssim 0.05$.  By separating the loss in fitting factor due to differences
in higher-order circular PN corrections and the effect of eccentricity, we conclude that current LIGO searches are
sensitive to binaries with small residual eccentricities when the waves enter
the sensitive band of the detector. However for eccentricities $e_0 \gtrsim 0.1$,
significant losses in sensitivity will be observed. 
For AdvLIGO, our results were dominated by differences in the PN order of the
signals and templates and so more investigation with higher-order PN eccentric
signals is needed. However, our study suggests that AdvLIGO is even more
sensitive to the effect of residual eccentricity in compact binary inspiral,
as would be expected.

In this study, we did not consider the ability of the LIGO detectors to
\emph{measure} eccentricity, however our results suggest that for $e_0 \gtrsim
0.1$, LIGO is sensitive to eccentricity, even in the current
detectors. Careful study of parameter estimation and implementation of a
search for eccentric systems with Enhanced LIGO or AdvLIGO would allow the
formation mechanisms proposed in Refs.~\cite{O'Leary:2008xt,Wen:2002km} to be
explored. 

\acknowledgments We thank Ian Hinder, Eric Poisson, and Deirdre
Shoemaker for helpful discussions on the subject of post-Newtonian modeling of 
eccentric binaries, Andrew Lundgren and Larne Pekowsky for their
contributions to the template bank simulation code and Sukanta Bose for
comments on this manuscript. This work was supported by
NSF grant PHY-0847611.  

\end{document}